\documentclass[paper]{geophysics}

\newcommand{\rs}[1]{\mathstrut\mbox{\scriptsize\rm #1}}

\newcommand{\diverg}{\nabla \cdot}
\newcommand{\ub}{\mathbf{u}}
\newcommand{\ddtt}{\partial^2_t}
\newcommand{\ddt}{\partial_t}

\newcommand{\f}{\mathbf{f}}
\newcommand{\x}{\mathbf{x}}
\newcommand{\xr}{\mathbf{x}_r}
\newcommand{\xs}{\mathbf{x}_s}

\newcommand{\wb}{\mathbf{w}}
\newcommand{\vb}{\mathbf{v}}
\newcommand{\s}{\mathbf{s}}
\newcommand{\Lb}{\mathbf{L}}
\newcommand{\I}{\mathbf{I}}
\newcommand{\Wr}{\mathbf{\Lambda}_r}

\newcommand{\nb}{\mathbf{n}}

\newcommand{\K}{\mathbf{K}}
\newcommand{\D}{\mathbf{D}}
\newcommand{\G}{\mathbf{G}}
\newcommand{\N}{\mathbf{N}_{\x}}
\newcommand{\matrixK}{\pmatrix{1 & 0 \cr 0 & -\mathbf{I}}}

\newcommand{\ddx}{\partial_x}
\newcommand{\ddz}{\partial_z}
\newcommand{\ddtau}{\partial_{\tau}}
\newcommand{\matrixN}{\pmatrix{0 & \nb^T \cr \nb & \mathbf{0}}}

\renewcommand{\figdir}{./}

\usepackage{tikz}
\usepackage{multirow}

\begin{document}

\title{Vector acoustic full waveform inversion: taking advantage of de-aliasing and receiver ghosts}

\renewcommand{\thefootnote}{\fnsymbol{footnote}} 


\address{
\footnotemark[1]Memorial University of Newfoundland \\
Department of Earth Sciences, \\
St. John's, NL A1C 5S7, Canada }
\author{Polina Zheglova\footnotemark[1] and Alison Malcolm\footnotemark[1]}

\footer{}
\righthead{Vector acoustic FWI: de-aliasing, receiver ghosts}

\maketitle

\begin{abstract}
Data acquisition is changing to incorporate more components of the recorded wavefield. An example of this is so-called vector data in which both the pressure and particle velocity are recorded in marine data. We present a vector acoustic full waveform inversion (VAFWI) method. We demonstrate the connection of the VAFWI adjoint operator with inverse wavefield extrapolation and show that under the assumption of a plane wave propagating towards an infinite flat recording surface at normal incidence, the VAFWI adjoint is equivalent to inverse extrapolation of the normal derivative of the recorded field. Thus, unlike the conventional FWI adjoint, the VAFWI adjoint is an inverse wavefield extrapolation operator. If these assumptions are violated, this equivalence relation is no longer true and becomes an approximation. We argue that this has implications in the handling of receiver ghosts by the two inversion methods: in the VAFWI adjoint, the receiver ghosts interact constructively with the back-propagated reflected field. We show numerically that the de-aliasing property of vector data extends to VAFWI, which results in fewer artefacts in VAFWI images compared to conventional FWI when data are spatially undersampled and aliased. Additional information about the subsurface contained in properly handled receiver ghosts can be utilized to achieve further VAFWI image improvement.
\end{abstract}

\section{Introduction}

Vector acoustic (VA) marine data, or dual-sensor data, consisting of hydrophone (pressure) and geophone (particle velocity or acceleration) measurements have been acquired and used for some time. Vector data contain directional information about the recorded wavefield, and have been utilized for a number of purposes, including wavefield separation and de-ghosting \citep{Carlson:2007,Reiser:2012}, sea surface wave-height estimation \citep{Orji:2012}, surface multiple suppression \citep{Soellner:2008}, data interpolation \citep{Robertsson:2008,Vassallo:2010,Ozbek:2010} and others. Reverse-time migration using vector data has also been proposed \citep{Fleury:2013,Ravasi:2015}. Such studies demonstrate the benefits of using vector data in the acoustic formulation in data processing and imaging workflows, compared to pressure data alone. Recently, interest has emerged to applying vector acoustic data in full waveform inversion (FWI) \citep{Akrami:2017,Zheglova:2018,Zheglova:2019,Zhong:2019,Kohnke:2019}. In this paper we attempt to answer the question: what kind of improvements from the directional character of vector data can be obtained in VAFWI and why?

The use of vector data is used for data interpolation, particularly in the cross-line direction in 3D, where the intervals between data samples can be quite large, resulting in spatial data aliasing. Horizontal aliasing leads to noisy images, where the noise in the image domain has the form of clutter \citep{Yilmaz:2008}. \citet{Robertsson:2008} show that if the cross-line component of particle velocity is available, the data can be accurately interpolated in the cross-line direction at half the Nyquist sampling rate. \citet{Vassallo:2010} develop an iterative matching pursuit algorithm that is able to reconstruct the missing data at above twice the Nyquist frequency. \citet{Ozbek:2010} generalize this method to simultaneous 3D reconstruction and de-ghosting. They show that using the vertical velocity component in addition to pressure and the horizontal (cross-line) velocity component improves the interpolation result. They conclude that the vertical velocity component carries significant information in the cross-line direction. \citet{Fleury:2013} in their vector acoustic reverse time migration (VARTM) study also demonstrate the reduction of the noise caused by data aliasing in the VA adjoint fields. In this paper, we show that the de-aliasing property of vector data carries over to VAFWI.

In the case of spatially undersampled data, receiver ghosts can potentially present a source of valuable additional information about the subsurface. In the context of conventional FWI, which utilizes only pressure data, especially for towed streamer marine acquisition, the downside of including receiver ghosts is that they have opposite polarity with respect to the up-going pressure waves. This results in destructive interference of the up-going field and its ghost at certain frequencies, including the useful low frequencies. Therefore, it is considered a desirable data pre-processing practice to suppress receiver ghosts. In the case of reverse time migration (RTM) of ocean bottom sensor data, where the loss of the low frequency content of the data may be less of an issue, receiver ghosts can cause spurious reflector artifacts in RTM images \citep{Ravasi:2015}. 

\citet{Ravasi:2015} propose to suppress these artifacts by combining pressure and calibrated velocity component data in the framework of wavefield reconstruction or inverse extrapolation. Their method relies on the fact that an acoustic wavefield can be exactly extrapolated into a subsurface region from measurements of pressure and particle velocity on the boundary of this region, if the Green's function inside the region is known. We have shown in our previous work \citep{Zheglova:2019} that the VAFWI adjoint operator handles receiver ghosts differently than the conventional FWI adjoint, which results in constructive rather than destructive interference of the up-going pressure wavefield and the receiver ghost in the adjoint field. In the case of ocean bottom cables, when receivers are located at a considerable depth, this also helps to avoid the spurious reflection events in the VAFWI gradients arising from cross-talk of various arrivals in the gradient, which is also numerically demonstrated by \cite{Zhong:2019}. In this paper we go one step further, and connect the VAFWI adjoint to inverse extrapolation. We show that under the specialized assumption of a plane wave propagating in a homogeneous earth to an infinite receiver plane at normal incidence, the VAFWI adjoint operator exactly inversely extrapolates the normal derivative of the incident field. When any of the above conditions are violated, the inverse extrapolation by the VAFWI adjoint is approximate. We clarify from this point of view the difference in handling of receiver ghosts by VAFWI and conventional FWI. We also show that for streamer acquisition, when the sensors are at a shallow depth below the water surface, this results in better preservation of the low frequencies in the VAFWI adjoint fields compared to FWI adjoint fields, if receiver ghosts are present in the data.

\citet{Meier:2010} show that vector data make no difference in the handling of the source ghost: instead of vector data at receivers, source de-ghosting requires directional (vector) sources. The main challenge in the creation of the directional sources lies in designing marine dipole source actuators capable of producing large enough energy at low frequencies that could be used in reflection seismic applications. Alternatively, \citet{Robertsson:2012} propose using closely spaced airgun arrays to generate separate shot gathers that can be subtracted to simulate a dipole source response. \citet{Fleury:2013} show that combinations of monopole and dipole sources result in directional sources, which together with vector data lead to better focusing of events on reflectors in RTM images. Closely spaced directional dipole sources do not seem to make a significant difference in VAFWI \citep{Akrami:2017}. In this paper we do not consider the effect of directional sources and focus solely on studying the effects of vector data on the receiver side. 

Multi-component data have been used in full waveform inversion in both the acoustic and elastic formulations in the time and frequency domain, including methods attempting to incorporate the advantages of both time and frequency domain formulations. Such inversion methods often incorporate nested hierarchical workflows \citep[][ and others]{Choi:2008, Brossier:2009, Asnaashari:2012, Plessix:2013, Prieux:2013, Yang:2014}. The possibility of poro-elastic FWI has also been considered \citep{Yang:2018, Yang:2019}. Despite its undeniable advantages, elastic FWI has not yet become a mainstream technology in industry due to its high computational cost. Acoustic inversion of vector data, on the other hand, does not require a significant increase in computational resources compared to conventional FWI but has certain advantages. It can be done in both time and frequency domain and hierarchical inversion strategies are also applicable. In some situations, formulations involving local solvers \citep{Kohnke:2019} can be used to reduce the computational cost. The disadvantage of the acoustic formulation is that the acoustic model describes wave propagation in the earth less accurately than the elastic model, resulting in a larger mismatch between the modelled and observed wavefields.

A time domain vector acoustic full waveform inversion was proposed by \citet{Akrami:2017} and generalized to multi-parameter case by \citet{Zheglova:2018}. \citet{Zhong:2019} introduce it in a different formulation with a source independent objective function and look at radiation patterns of each data component. While these authors emphasize the difference in the VAFWI and conventional FWI adjoints, they do not analyze this in detail. \cite{Kohnke:2019} apply the VAFWI in a cross-borehole synthetic study with a local domain. 

In this paper, we formulate the method in the time domain. We then connect the VAFWI adjoint operator to inverse wavefield extrapolation. We show that unlike the conventional FWI adjoint, the VAFWI adjoint becomes an inverse extrapolation operator under certain specialized conditions. From this point of view we clarify the difference in the handling of the receiver ghost fields by VAFWI and conventional FWI adjoint operators. Then we demonstrate numerically the advantages of VAFWI over conventional acoustic FWI in the case of spatially undersampled data with and without the free surface. Finally, we draw conclusions and outline possible future research directions.



\section{Method formulation}

Just as in conventional FWI, VAFWI assumes that the seismic wave propagation in the subsurface is governed by the acoustic wave equation. Thus, only the P-wave part of the wavefield is used in the inversion.

The methodology for mono- and multiparameter VAFWI is first introduced in our earlier work \citep{Akrami:2017,Zheglova:2018}. Our formulation follows closely that of \citet{Fleury:2013}: it is a pressure-displacement time domain formulation. We summarize the method in this paper for completeness. 

\subsection{The forward problem}

The forward problem is given by the variable density vector acoustic wave equation:
\begin{eqnarray}
	\Lb \wb(\x,\xs,t) = \s(\x,t) \nonumber\\
	\wb(\x,\xs,0) = \ddt \wb(\x,\xs,0) = 0
	\label{forward_matrix}
\end{eqnarray}
where $\Lb$, $\wb$ and $\s$ are the vector acoustic forward modelling, vector data and vector source respectively given by:
\begin{equation}
	\Lb = \pmatrix{ \frac{1}{\rho c^2} & \nabla^T \cr \nabla & \rho \ddtt \mathbf{I}}, \;\;
	\wb = \pmatrix{p \cr \ub}, \;\;
	\s = 	\pmatrix{\frac{1}{\rho c^2}q \cr \f }.
	\label{VAO}
\end{equation}
Here $\nabla^T$ (dotted with $\ub$) is the divergence operator, 
	$\I$ is the $n \times n$ identity matrix (in $n$ dimensions)
and the rest of the variables/parameters are defined in Table~\ref{tbl:notation}.

\tabl{notation}{List of notations with units (where applicable).}{
  	\begin{center}
    	\begin{tabular}{|c|l|c|}
      		\hline
      		  	Notation & Name & Units (3D) \\
		\hline
      		\hline
     			$p$ & pressure & $\frac{\mbox{N}}{\mbox{m}^2} = \frac{\mbox{kg}}{\mbox{m s}^2}$  \\
		\hline
			$\ub$ & particle displacement & m \\
		\hline 
			$\wb = \pmatrix{p & \ub}^T$ & vector data & \\
		\hline
			$c$ & P-wave velocity of the subsurface & m/s \\
		\hline
			$\rho$ & density of the subsurface & $\frac{\mbox{kg}}{\mbox{m}^3}$ \\
		\hline
			$\s = \pmatrix{q & \f}^T$ & vector source & \\
		\hline
			$q$ & monopole (volume injection) source & $\frac{\mbox{kg}}{\mbox{m s}^2}$ \\
		\hline
			$\f$ & dipole (point force) source &  $\frac{\mbox{kg}}{\mbox{m}^2 \mbox{s}^2}$ \\
		\hline
			$\x$ & spatial variable & m \\
		\hline
			$t$ & time variable & s \\
		\hline
			$\xs$ & source locations & m \\
		\hline 
			$\xr$ & receiver locations & m \\
		\hline
			$s$, $r$ & source and receiver counters & \\
		\hline
			$T: t \in [0,T]$ & record length & s \\
		\hline
			$\wb^{rec} = \pmatrix{p^{rec} & \ub^{rec}}^T$ & recorded vector data &\\
		\hline
    	\end{tabular}
  	\end{center}
}

A minor deviation of our formulation from that of \citet{Fleury:2013} is the scaling of the first equation in the vector acoustic system: the model parameters scale the time derivative instead of the divergence. With this formulation, the adjoint solver is easily obtained from the forward solver. Moreover, time-derivatives of the displacement field are used in the gradient calculation instead of the spatial derivatives, and no spatial interpolation of the fields is required. We still require time derivatives and potentially interpolation in time, but the time axis is normally finely sampled for stability reasons, so our formulation modification leads to a more accurate gradient calculation, as verified by adjoint tests. 

If density is assumed constant, we make a change of variables: $\rho \ub \mapsto \ub$, and matrix-vector quantities in (\ref{forward_matrix}) become:
\begin{eqnarray}
	\Lb = \pmatrix{m & \nabla^T \cr \nabla & \ddtt \mathbf{I} }, \;\;
	\wb = \pmatrix{ p \cr \ub }, \;\;
	\s = \pmatrix{ m q \cr \f },
	\label{VAO_cd}
\end{eqnarray}
where $m = 1/c^2$ is the squared slowness.

\subsection{Objective function}

The objective function we use is:
\begin{equation}
	J = \frac{1}{2} \sum_{s,r} \int_0^T \| \Wr[\wb(\xr,\xs,t)-\wb^{rec}(\xr,\xs,t)] \|^{2} \; dt,
	\label{obj}
\end{equation}
where 
$\Wr$ is a data weighting operator that ensures that the contributions from the different data components are balanced in amplitude and have the same physical units, and the rest of the variables are defined in Table~\ref{tbl:notation}. 
For variable density VA operator, we take the data weighting operator proposed by \cite{Fleury:2013}, equation (22):
\begin{equation}
	\Wr = \pmatrix{ \frac{1}{\sqrt{\rho c^2}} & \mathbf{0}^T \cr
		\mathbf{0} & \sqrt{\rho}\ddt \mathbf{I}}.
\end{equation}
If density is assumed constant, with redefinition of $\ub$ described above the weighting operator becomes 
\begin{equation}
	\Wr = \pmatrix{ \sqrt{m} & \mathbf{0}^T \cr \mathbf{0} & \ddt \mathbf{I}}.
\end{equation}

\subsection{Vector acoustic FWI gradient}

The gradient for the objective function (\ref{obj}) is calculated by the adjoint state method \citep{Plessix:2006, Fichtner:2011}. Since the application of the adjoint state method to inverse problems is well explained in these references, and equations similar to ours have been also derived by \cite{Fleury:2013}, we only show the final result. The adjoint problem for our inverse problem is:
\begin{eqnarray}
	{\Lb}^{\dagger} \wb^{\dagger}(\x,\xs,t) = \s^{\dagger}(\x,\xs,t) \label{adjoint_matrix}\\
	\wb^{\dagger}(\x,\xs,0) = \ddt \wb^{\dagger}(\x,\xs,0) = 0.
	\label{end_cond}
\end{eqnarray}
The adjoint operator ${\Lb}^{\dagger}$ is given by:
\begin{eqnarray}
	{\Lb}^{\dagger} = \pmatrix{ \frac{1}{\rho c^2} & - \nabla^T \cr 
		-\nabla & \rho \ddtt \mathbf{I} }
\label{VAO_adjoint}
\end{eqnarray}
and the adjoint sources are given by:
\begin{equation}
	\s^{\dagger}(\x,\xs,t) = -\sum_r \Wr^{\dagger} \Wr [\wb(\xr,\xs,T-t)-\wb^{rec}(\xr,\xs,T-t)] \delta(\x-\xr), 
	\label{adj_src}
\end{equation}
where $\Wr^{\dagger}$ is the adjoint of $\Wr$:
\begin{equation}
	\Wr^{\dagger} = \pmatrix{ \frac{1}{\sqrt{\rho c^2}} & \mathbf{0}^T \cr
		\mathbf{0} & - \sqrt{\rho}\ddt \mathbf{I}}.
	\label{Lambda_adj}
\end{equation}

Assuming that the subsurface parameters to be inverted for are velocity and density $(c, \rho)$, the gradient of the objective function with respect to them is computed by:
\begin{eqnarray}
	\frac{\partial J}{\partial c} &=&  \sum_s \int_0^T \left( - \frac{2}{\rho c^3} \; p^{\dagger}(\x,\xs,T-t) p(\x,\xs,t) \right) \;  dt \\
	\frac{\partial J}{\partial \rho} &=& \sum_s \int_0^T \left( - \frac{1}{\rho^2 c^2} \; p^{\dagger}(\x,\xs,T-t)  p(\x,\xs,t)  + \ub^{\dagger}(\x,\xs,T-t)  \cdot  \ddtt \ub(\x,\xs,t) \right)  \; dt.
	\label{crho}		
\end{eqnarray}

If the density is assumed constant, it is a convenient and widely used practice to invert for squared slowness $m$. In this case, in the adjoint equation, like in the forward modelling equations, the factors $\frac{1}{\rho c^2}$ and $\rho$ are replaced by $m$ and 1 respectively. The gradient with respect to $m$ is computed by:
\begin{equation}
	\frac{\partial J}{\partial m} =  \sum_s \int_0^T p^{\dagger}(\x,\xs,T-t) \; p(\x,\xs,t) \; dt.
	\label{grad_m}
\end{equation}
We  note that even though the gradient in equation (\ref{grad_m}) depends directly only on the modelled and adjoint pressure, it indirectly depends on both pressure and displacement data through the adjoint sources, equation (\ref{adj_src}). In the case of variable density, other pairs of parameters can also be recovered, e.g. velocity $c$ and acoustic impedance $I_p$, squared slowness $m$ and density $\rho$, or else, compressibility $\kappa = \frac{1}{\rho c^2}$ and density $\rho$, however, apart from some remarks, comparison of parametrization effects is outside the scope of this paper.

We solve equations (\ref{forward_matrix}) and (\ref{adjoint_matrix}) using a pressure-particle velocity $p$-$\vb$ acoustic formulation on a staggered grid in space and time. Displacement is calculated on the fly. We apply perfectly matched layer (PML) absorbing boundary conditions either on all sides of the computational domain or on all sides except the top boundary, where we also use the free surface boundary condition.

We solve the inverse problem iteratively, starting from the initial guess $c^{(0)}$ and $\rho^{(0)}$ (alternatively $m^{(0)}$ for the constant density case). The descent direction and step size are computed by the L-BFGS optimization method with line-search \citep{Nocedal:2006}.

In the examples below, we compare VAFWI to conventional FWI of only pressure data. We use an equivalent of the conventional FWI, obtained from our VAFWI formulation by setting everywhere the data weights to
\[
	\Wr =  \pmatrix{ 1 & \mathbf{0}^T \cr \mathbf{0} & 0 \cdot \mathbf{I}}.
	\label{Lambda}
\]
Thus, we present a unified VAFWI / FWI approach to inversion.

\subsection{Connection with inverse wavefield extrapolation}

If measurements of a vector acoustic field are available everywhere on a surface surrounding a region, and the Green's matrix for that region is known, then the field can be exactly extrapolated inside the region from the boundary measurements. This is done via Helmholtz-Kirchhoff formulas \cite[e.g.][]{Morse:1953, Aki:2002}. A question arises about the connection of wavefield extrapolation with VAFWI. \cite{Wapenaar:2004} and \cite{Wapenaar:2007} derive a general form of extrapolation equations for the first order matrix-vector system of partial differential equations describing a range of wave propagation, flow and diffusion phenomena, of which vector acoustics is a special case. Following their notation, in this section we connect the VAFWI adjoint operator with the inverse wavefield extrapolation \cite[][]{Wapenaar:2007} and show that under certain specialized assumptions, the VAFWI adjoint operator becomes an inverse wavefield extrapolation operator. 

Inverse wavefield extrapolation is schematically illustrated in Figure~\ref{fig:inv_ext}. In this Figure, the space shown as a 2D plane is divided into two subdomains, $D$ and $D'$, separated by a recording surface $\partial D$. This surface need not be a real boundary and can be assumed transparent. In inverse extrapolation, the wavefield $\wb=\pmatrix{p & \ub}^T$ incident upon the surface $\partial D$ is extrapolated in reverse time order into $D$ from its measurements everywhere on $\partial D$. In Figure~\ref{fig:inv_ext}, $D$ is compact and $\partial D$ is closed, however, similar to a standard full waveform inversion configuration, $D$ can be an infinite subdomain, i.e. a half-space. In this case, the part of $\partial D$ plotted in dashed lines does not contribute to the extrapolation, and the part of $\partial D$ plotted with a solid line extends to infinity. The field $\wb$ is extrapolated from the part of $\partial D$ plotted as a solid line. 

\plot{inv_ext}{trim = 5cm 13cm 7cm 8cm, clip=true,width=0.6\textwidth}
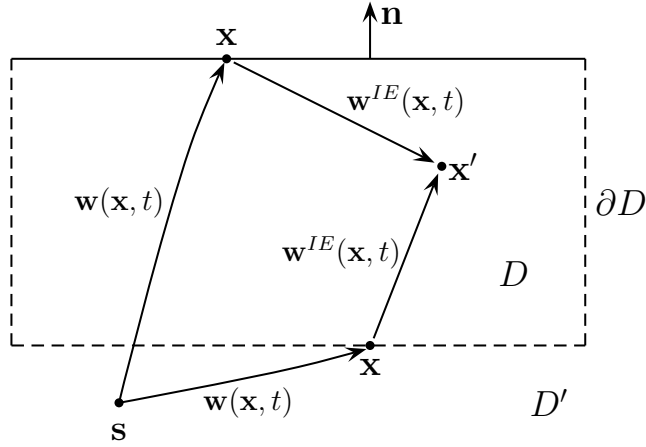
{Schematic of inverse extrapolation. The field $\wb$ propagates from the source $\s$ (assumed in this figure outside $D$) towards points $\x \in \partial D$ where it is recorded. Then the field is extrapolated in reverse-time order from $\partial D$ to points $\x' \in D$.} 

In order to facilitate connection of our adjoint operator with the inverse extrapolation operator, we recast our forward problem as a first order system of equations. This is done by introducing a new variable: $\tilde{\wb} = \pmatrix{p & \vb}^T$, where $\vb = \ddt \ub$ is particle velocity. Our forward problem in terms of the new variable is:
\begin{equation}
	\tilde{\Lb} \tilde{\wb} = \tilde{\s}
	\label{ltilde}
\end{equation}
where
\[
	\tilde{\Lb} = \pmatrix{ \frac{1}{\rho c^2}\ddt & \nabla^T \cr \nabla & \rho \ddt \mathbf{I}}, \;\;
	\tilde{\s} = \pmatrix{\frac{1}{\rho c^2} \ddt q \cr \f }.
\]
Throughout this section quantities marked by the tilde will be associated with this 1$^{st}$ order system formulation.

The inverse extrapolation is performed according to equation (67) of \citet{Wapenaar:2007} formulated in the frequency domain, from which we obtain the following corresponding equation in the time domain:
\begin{equation}
	\tilde{\wb}^{IE}(\x,t) = \int\limits_{\partial D} (\K \tilde{\G}(\x,\x',t) \K) *_t \mathbf{N}_{\x'} \tilde{\wb}(\x',T-t) \; dS',
	\label{wr_int}
\end{equation}
where 
\begin{description}
	\item $\tilde{\wb}^{IE}(\x,t)$ is the inversely extrapolated field such that
		\begin{equation}
			\tilde{\wb}^{IE}(\x,t) = 
				\left\{ \begin{array}{l}
					\tilde{\wb}(\x,T-t), \;\;\; \x \in D \\
					0, \;\;\; \;\;\; \;\;\;\;\;\; \;\;\; \;\;\; \; \x \in D',	
				\end{array}
			\right.
		\end{equation}
	\item $\K$ is a diagonal matrix given by:
		\[ \K = \matrixK, \]
	\item $\tilde{\G}(\x,\x',t)$ is the Green's matrix corresponding to the forward operator $\tilde{\Lb}$ 
	\item $\N$ is a matrix of outward normal components $\nb$ to $\partial D$ organized in the same order as the spatial derivatives in the operators $\Lb$ and $\tilde{\Lb}$:
		\[ \N = \matrixN .\]
\end{description}
We multiply equation (\ref{wr_int}) by $\K$ on both sides, and rearrange the integrand to obtain:
\begin{equation}
	\K \tilde{\wb}^{IE}(\x,t) = \int\limits_{\partial D} \tilde{\G}(\x,\x',t)  *_t \K \mathbf{N}_{\x'} \tilde{\wb}(\x',T-t) \; dS',
	\label{wr_int1}
\end{equation}
which is equivalent to solving the following system:
\begin{equation}
	\tilde{\Lb}(\x,t) \K \tilde{\wb}^{IE}(\x,t) = \K \mathbf{N}_{\x} \tilde{\wb}(\x,T-t) \delta_{\partial D},
	\label{wr_int1}
\end{equation}
where $\delta_{\partial D}$ represents a surface source along $\partial D$. When written component-wise, equation (\ref{wr_int1}) becomes:
\begin{eqnarray}
	\frac{1}{\rho c^2} \ddt p^{IE}(\x,t) - \diverg \vb^{IE}(\x,t) &=& \nb \cdot \vb(\x,T-t)  \delta_{\partial D} \nonumber\\
	\rho \ddt \vb^{IE}(\x,t) -\nabla p^{IE}(\x,t)  &=& \nb \; p(\x,T-t)  \; \delta_{\partial D}.
	\label{tilde_eq}
\end{eqnarray}
 Alternatively, we can make a change of variables $\tau = T-t$ and rewrite system (\ref{tilde_eq}) in terms of the reverse time $\tau$ as:
\begin{eqnarray}
	- \frac{1}{\rho c^2} \ddtau p^{IE}(\x,T-\tau) - \diverg \vb^{IE}(\x,T-\tau) &=& \nb \cdot \vb(\x,\tau)  \delta_{\partial D} \nonumber\\
	- \rho \ddtau \vb^{IE}(\x,T-\tau) -\nabla p^{IE}(\x,T-\tau)  &=& \nb \; p(\x,\tau)  \; \delta_{\partial D}
	\label{tilde_eq1}
\end{eqnarray}
or
\begin{equation}
	\tilde{\Lb}^{\dagger}(\x,\tau) \tilde{\wb}^{IE}(\x,T-\tau) = \tilde{\s}^{I.E.}(\x,\tau)  \;\;\;\;\; \forall \x \in D
	\label{wr_diff}
\end{equation}
where 
\[
	\tilde{\Lb}^{\dagger}(\x,\tau) = \pmatrix{ -\frac{1}{\rho c^2}\partial_{\tau} & -\nabla^T \cr -\nabla & -\rho \partial_{\tau} \mathbf{I}}
\]
is the formal adjoint of $\tilde{\Lb}$. Thus, we have expressed the inverse extrapolation integral (\ref{wr_int}) as a differential equation system involving the adjoint of the forward modelling operator. 

Returning now to pressure / displacement formulation, we note that in equation (\ref{tilde_eq}), $\tilde{\wb}^{IE}$ is calculated forward in time, therefore we postulate the following relationship between the inversely extrapolated velocity and displacement: $\vb^{IE} = \ddt \ub^{IE}$, while on the right hand side of the first equation in (\ref{tilde_eq}) $\vb(\x,T-t) = -\ddt \ub(\x,T-t)$. Then, integrating the first equation of (\ref{tilde_eq}) with respect to $t$, we obtain
\begin{eqnarray}
	\frac{1}{\rho c^2} p^{IE}(\x,t) - \diverg \ub^{IE}(\x,t) &=& - \nb \cdot \ub (\x,T-t) \delta_{\partial D} \nonumber\\
	\rho \ddtt \ub^{IE}(\x,t) - \nabla p^{IE}(\x,t)  &=& \nb \;p(\x,T-t) \; \delta_{\partial D}.
	\label{pu_eq}
\end{eqnarray}
We can write equation (\ref{pu_eq}) in matrix form as:
\begin{equation}
	{\Lb}^{\dagger}(\x,t) \wb^{IE}(\x,t) = \s^{I.E.}(\x,t).
	\label{wr_diff1}
\end{equation}
where ${\Lb}^{\dagger}$ is our VAFWI adjoint operator given by equation (\ref{VAO_adjoint}), and the source is given by
\begin{equation}
	\s^{I.E.}(\x,t) = \pmatrix{ -\nb \cdot \ub \cr \nb p }(\x,T-t)\; \delta_{\partial D},
	\label{IEsource}
\end{equation}
Alternatively, by making a change of variables $\tau = T-t$, we can rewrite equation (\ref{pu_eq}) using the reverse time variable $\tau$ analogous to (\ref{tilde_eq1}) as
\begin{eqnarray}
	\frac{1}{\rho c^2} p^{IE}(\x,T-\tau) - \diverg \ub^{IE}(\x,T-\tau) &=& - \nb \cdot \ub (\x,\tau) \delta_{\partial D} \nonumber\\
	\rho \ddtt \ub^{IE}(\x,T-\tau) - \nabla p^{IE}(\x,T-\tau)  &=& \nb \;p(\x,\tau) \; \delta_{\partial D}.
	\label{pu_eq1}
\end{eqnarray}
Thus, essentially, the inverse extrapolation amounts to back-propagation of the normal components of the incident field from $\partial D$ with the help of the adjoint operator ${\Lb}^{\dagger}$. The resulting field $\wb^{IE}$ is such that 
\begin{equation}
	\wb^{IE}(\x,t) = 
		\left\{ \begin{array}{l}
			\wb(\x,T-t), \;\;\; \x \in D \\
			0, \;\;\; \;\;\; \;\;\;\;\;\; \;\;\; \;\;\; \; \x \in D'.	
		\end{array}
	\right.
\end{equation}

Consider now the application of the VAFWI adjoint problem (\ref{adjoint_matrix}), (\ref{end_cond}) to the wavefield $\wb$. That is, we treat $\wb$, rather than the data residual, as the input data to the VAFWI adjoint problem, where instead of isolated receiver locations we assume a continuous receiver surface. From (\ref{adj_src}) and (\ref{Lambda_adj}) 
\begin{eqnarray}
	\s^{\dagger}(\x,t) =-
		\Wr^{\dagger} \Wr \wb \; \delta_{\partial D} = 
		\pmatrix{ - \frac{1}{\rho c^2} p \cr \rho \ddtt \ub}(\xr,\xs,T-t) \; \delta_{\partial D},
		\label{sadj}
\end{eqnarray}
where in obtaining the adjoint sources, we first perform time differentiation and then time-reversal.  Using (\ref{forward_matrix}), (\ref{VAO}), we can rewrite the VAFWI adjoint source as:
\begin{eqnarray}
	\s^{\dagger}(\x,t) =
 		\pmatrix{ 
		 \diverg \ub  \cr
		- \nabla p} (\xr,\xs,T-t) \; \delta_{\partial D}.
		\label{sadj1}
\end{eqnarray}
We note that (\ref{sadj1}) differs from (\ref{IEsource}) by a change of sign and a replacement of the normal components by the corresponding spatial derivative components of the recorded fields.

Retracing the steps that lead from (\ref{wr_int}) to (\ref{pu_eq}), we conclude that we can compactly represent our VAFWI adjoint operator in the integral notation of (\ref{wr_int}) as
\begin{equation}
	\tilde{\wb}^{\dagger}(\x,t) = - \int\limits_{\partial D} (\K \tilde{\G}(\x,\x',t) \K) *_t \D_{\x'} \tilde{\wb} (\x',T-t) \; dS',
	\label{adj_int}
\end{equation}
where $\tilde{\wb}^{\dagger} = \pmatrix{p^{\dagger} & \vb^{\dagger}}^T$ with $\vb^{\dagger} = \ddt \ub^{\dagger}$, and $\D_{\x}$ is the spatial derivative part of the operator $\Lb$ (or $\tilde{\Lb}$):
\[
	\D_{\x} = \pmatrix{0 & \nabla^T \cr \nabla &\mathbf{0}}.
\]
That is, our adjoint operator injects on $\partial D$ the spatial derivative components of the recorded wavefields, rather than the corresponding normal components of the those wavefields, as would be the case in the inverse extrapolation. In general, the VAFWI adjoint is not an inverse extrapolation operator, as we show in the following section. That is, in general, it does not equal some time-reversed wavefield inside $D$, and it does not vanish in $D'$, except under certain conditions, which we consider in the next section. If those conditions are approximately satisfied, the the VAFWI adjoint approximates an inverse extrapolation operator. This has implications for VAFWI, which we consider also in the following section.

{\bf Remark.} 
Formula (\ref{wr_int}) holds under the assumption that no sources of $\wb$ are present inside $D$. If sources are present inside this region, then a sink needs to be added to the above formula to cancel the sources and correctly extrapolate the field \citep{Cassereau:1992}. The sink term can be derived from the correlation reciprocity relation \citep{Wapenaar:2007}.

\subsubsection{Implications for VAFWI}

The implications of equation (\ref{adj_int}) for VAFWI gradient calculation become most apparent if we apply the VAFWI adjoint operator to the records of a plane wave $\wb$ propagating in a homogeneous background model at normal incidence to a flat infinite recording surface $\partial D$. To be specific, we assume that $\partial D$ is horizontal, i.e. $\partial D = \{\x =(x,z): z=z_0\}$, where $x$ is the horizontal coordinate, and $z$ is the vertical (depth) coordinate oriented downward, $D = \{(x,z): z>z_0\}$ and $D' =  \{(x,z): z<z_0\}$. In this case, $\nb = \pmatrix{0 & -1}^T$, the horizontal displacement and the horizontal derivative of pressure are zero: $u_x  = \ddx p = 0$, and 
\[
	\s^{\dagger} = \pmatrix{\ddz u_z \cr 0 \cr -\ddz p} =  \ddz \pmatrix{u_z \cr 0 \cr -p}
	= \ddz \pmatrix{ \nb \cdot \ub \cr \nb p } = \ddz \s^{I.E.}.
\] 
Thus, our adjoint operator inversely extrapolates the vertical spatial derivative of the incident field.

We illustrate this on a 2D example in Figure \ref{fig:pz_t=0_3s,p_adj_VA_t=0_3s,p_adj_FWI_t=0_3s,pz_t=0_4s,p_adj_VA_t=0_4s,p_adj_FWI_t=0_4s}. The first column of images (\ref{fig:pz_t=0_3s} and \ref{fig:pz_t=0_4s}) represents time snapshots at time $t=0.3$ s and $t=0.4$ s of the $z$-derivative of a plane wave propagating up. The plane wave is generated by a line source located at depth $z=0.4$ km (red line), the peak frequency of the source wavelet is 10 Hz. Only pressure is shown in these images. The vector field (pressure and displacement) is recorded at the receiver surface located at the depth $z_0=0.1$ km (black line). We back-propagate these recorded data from $\partial D$ to the original times ($t=0.4$ and 0.3 s) using our VAFWI adjoint operator. The second column of images (\ref{fig:p_adj_VA_t=0_3s} and \ref{fig:p_adj_VA_t=0_4s}) shows snap-shots of the resulting adjoint field at these times (only adjoint pressure is shown). We see that the adjoint field is extrapolated only in the domain $D$:  $z>0.1 \mbox{ km}$, where it equals the $z$-derivative of the original field, and it is zero in $D'$. Thus our VAFWI adjoint field represents an inverse-extrapolation operator under these specialized conditions. The last column of images (\ref{fig:p_adj_FWI_t=0_3s} and \ref{fig:p_adj_FWI_t=0_4s}) shows the snap-shots of the adjoint pressure field generated from the same data by the adjoint operator from conventional FWI. If we compare them with images \ref{fig:p_adj_VA_t=0_3s} and \ref{fig:p_adj_VA_t=0_4s}, we see that the  FWI adjoint is not an inverse extrapolation operator, since it back-propagates the recorded data both in $D$ and $D'$, and the adjoint field no longer equals $\ddz p$ in $D$.

\multiplot{6}{pz_t=0_3s,p_adj_VA_t=0_3s,p_adj_FWI_t=0_3s,pz_t=0_4s,p_adj_VA_t=0_4s,p_adj_FWI_t=0_4s}{width=0.3\textwidth,trim = 1cm 8.5cm 1cm 8.5cm, clip=true}
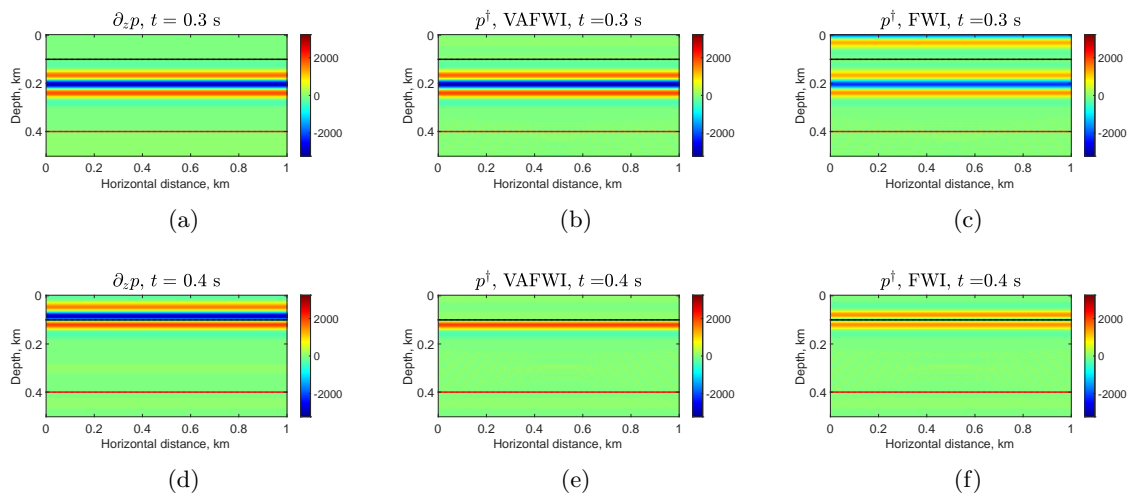
{(a), (d) Snapshots at times $t=0.3$ s and $t=0.4$ s of the $z$-derivative of a plane wave propagating up from a line source at depth $z=0.4$ km, and recorded at the recording surface located at  depth $z=0.1$ km. Only the pressure field $p$ is shown. (b), (e) Snapshots of the VAFWI adjoint pressure field $p^{\dagger}$ generated from the recorded data. (c), (f) Snapshots of the conventional FWI adjoint pressure field $p^{\dagger}$ generated from the same recorded data.}

If the velocity and density in $D$ are not constant, the boundary $\partial D$ is not flat, or the incoming wave is not a plane wave propagating to the recording surface at normal incidence angle, the VAFWI adjoint no longer extrapolates the vertical derivative of the incoming field exactly. Figure \ref{fig:dzp0_t=_25_close_up} shows snapshots of the $z$-derivative of the forward field from a point source located at depth $z=0.9$ km and horizontal distance $x=0.5$ km. Only the pressure field is shown. Figure \ref{fig:padj_t=_25_close_up_vfwi} shows the VAFWI adjoint pressure field generated from the records of this forward field at the recording surface at depth $z_0=0.1$ km. Most of the incoming energy is still extrapolated by VAFWI adjoint into $D$ and the extrapolated field approximates  $\ddz p$ at low incidence angles. The approximation gets worse, as the incidence angle increases. The artifacts on the sides of the model are due to the finite aperture of the recording surface  (truncation of $\partial D$) in our numerical simulation. The FWI adjoint pressure from the same data is shown in Figure \ref{fig:padj_t=_25_close_up_fwi} for comparison. The FWI adjoint field propagates in both $D$ and $D'$ in equal shares. In the presence of the free surface, the upgoing part of FWI adjoint field (as in Figures \ref{fig:p_adj_FWI_t=0_4s}, \ref{fig:p_adj_FWI_t=0_3s} and \ref{fig:padj_t=_25_close_up_fwi}) would be reflected from the free surface and generate spurious reflectors in the gradient update, in contrast to VAFWI.

\multiplot{3}{dzp0_t=_25_close_up,padj_t=_25_close_up_vfwi,padj_t=_25_close_up_fwi}{width=0.3\textwidth,trim = 1cm 8.5cm 1cm 8.5cm, clip=true}
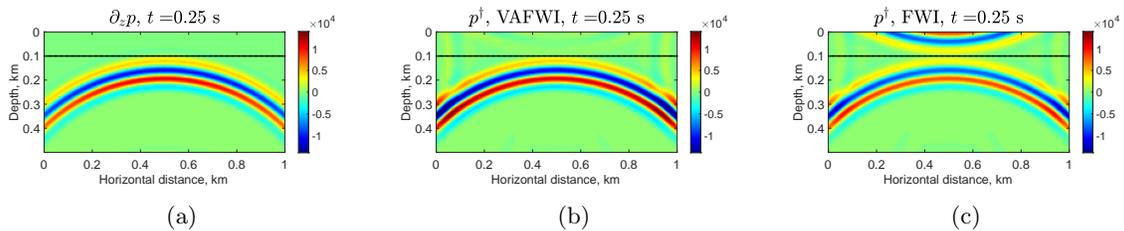
{(a) Snapshot at times $t=0.25$ s of the $z$-derivative of a circular wave propagating up from a point source at $(x,z)=(0.5,0.9)$ km and recorded at the recording surface located at depth $z=0.1$ km. Only the pressure field $p$ is shown. (b) Snapshot of the VAFWI adjoint pressure field $p^{\dagger}$ generated from the recorded data. (c) Snapshot of the conventional FWI adjoint pressure field $p^{\dagger}$ generated from the same recorded data. The artifacts at the sides are caused by truncation of the infinite boundary $\partial D$ in our numerical simulation.}

 \cite{Zheglova:2019} show that VAFWI and FWI adjoint operators handle receiver ghosts differently: VAFWI adjoint back-propagates receiver ghosts towards the free surface, where they reflect and change polarity, and then constructively interfere with the adjoint reflected field. Thus, VAFWI naturally separates the reflected arrival from the receiver ghost. By contrast, the FWI adjoint back-propagates both up-going reflected waves and the receiver ghosts in equal shares in both up and down directions, which results in cross-talk of these arrivals in the adjoint field and the gradient, leading to spurious events in the gradient. 

\multiplot{4}{padj_vfwi_z05,padj_fwi_z05,VAFWI_freq_spect,FWI_freq_spect}{width=0.45\textwidth,trim = 1cm 6.5cm 2cm 6.5cm, clip=true}
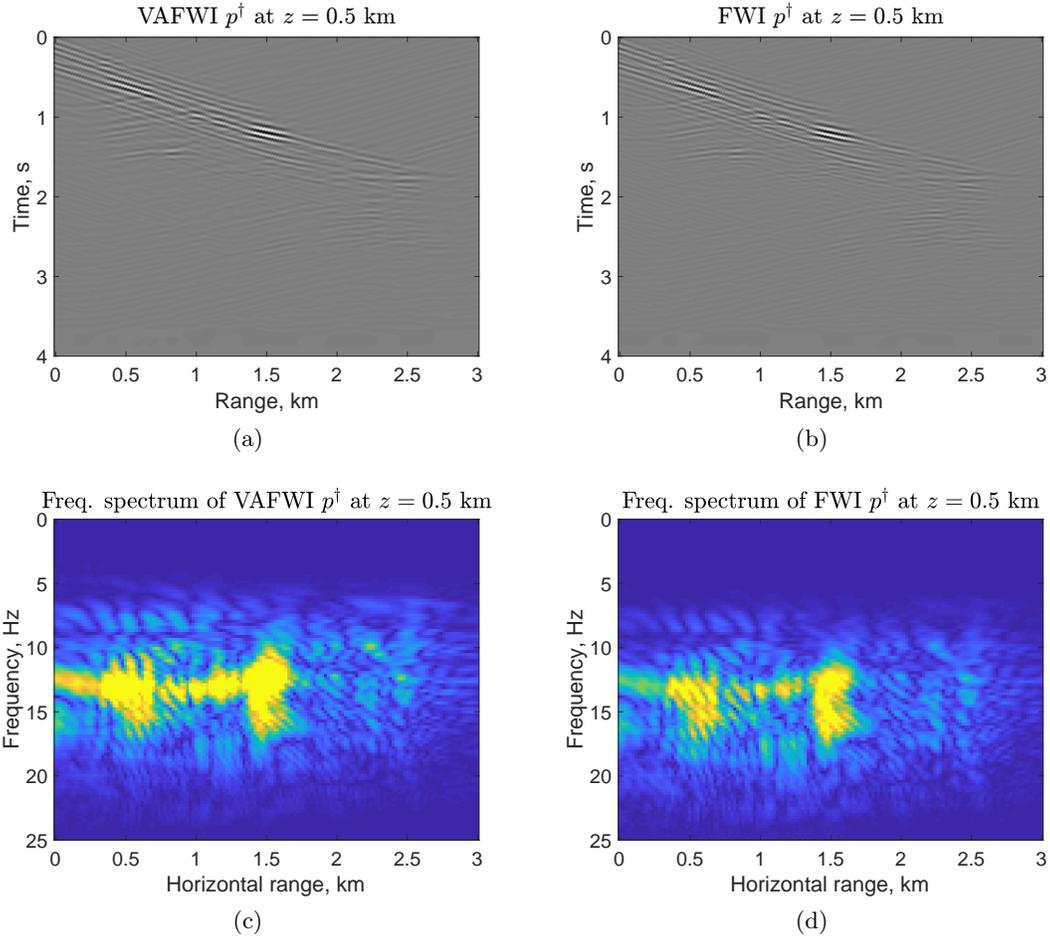
{(a), (b) Adjoint pressure fields generated by VAFWI and FWI from data residuals computed using modified truncated Marmousi-II model and an initial model obtained from it by Gaussian smoothing (Figure~\ref{fig:C_geop,C0_geop}). We show slices of the adjoint fields at the depth $z = 0.5$ km below the free surface for all times and horizontal distances. (c), (d) Frequency spectra of the above adjoint field slices. The spectra are plotted on the same scale, dark blue corresponds to 0 and bright yellow corresponds to $\max \left|F.F.T.\left(p^{\dagger}_{VAFWI}\right) \right|/2$. The seismic source for this synthetic experiment was placed at $x = 0.1$ km, $z=0.02$ km, receivers were placed at the same depth between $x = 0.1$ and $x = 2.9$ km every 20 m. Notice the broadening and the shift towards the low frequencies of the spectrum of the VAFWI adjoint field.}

\multiplot{2}{ampl_spec_x05_z05,ampl_spec_x1_z05}{width=0.45\textwidth,trim = 1cm 6.5cm 2cm 6.5cm, clip=true}
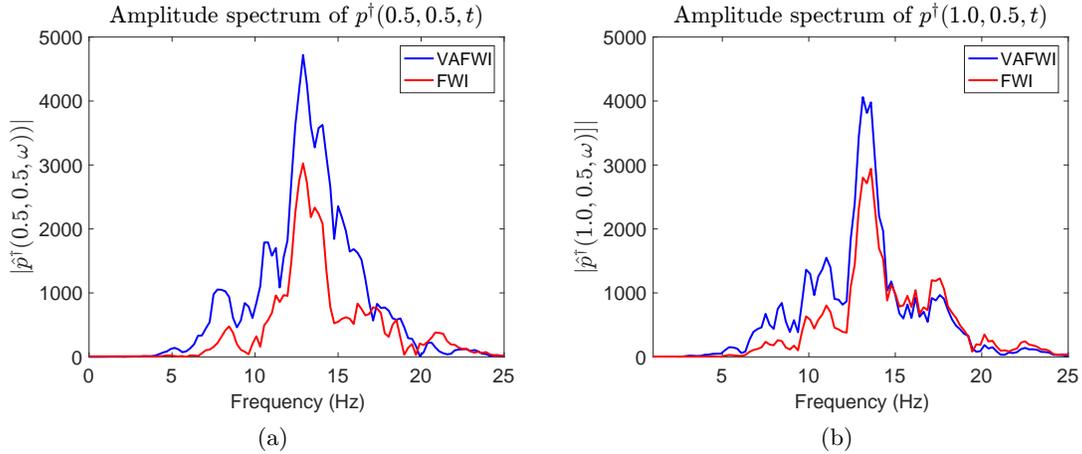
{(a), (b) Traces at $x=0.5$ km and $x=1$ km from the spectra shown in Figures \ref{fig:VAFWI_freq_spect} and \ref{fig:FWI_freq_spect}.}

For low-depth towed streamer marine acquisition, the separation of the receiver ghosts from the reflected arrival by VAFWI adjoint operation can also result in better preservation of the low frequency content in the data. In our experience, this effect is less pronounced with deeper receiver placement. Figures \ref{fig:padj_vfwi_z05} and \ref{fig:padj_fwi_z05} show the adjoint pressure fields generated by VAFWI and FWI adjoint operators from data residuals computed in the modified true and starting Marmousi-II P-wave velocity models shown in Figure \ref{fig:C_geop,C0_geop} (density is assumed constant in this experiment). The source for this simulation is placed at $x=0.1$ km in the horizontal direction at depth $z=0.02$ km below the surface, the source wavelet is a Ricker wavelet with peak frequency of 7 Hz. The receivers are placed at the same depth at even distances of 20 m between $x=0.1$ and $x=2.9$ km. The free-surface boundary condition is applied at the top of the model for data generation, modelling and adjoint calculation. In Figures \ref{fig:padj_vfwi_z05} and \ref{fig:padj_fwi_z05}, we show slices of the VAFWI and FWI adjoint pressure fields $p^{\dagger}$ at depth $z=0.5$ km for all times and horizontal distances. Figures \ref{fig:VAFWI_freq_spect} and \ref{fig:FWI_freq_spect} show the frequency spectra of the above field slices, and Figures \ref{fig:ampl_spec_x05_z05} and \ref{fig:ampl_spec_x1_z05} show plots of these spectra at two horizontal distances: $x=0.5$ and $x=1.0$ km. We observe a difference in the frequency content of the adjoint fields: particularly, low frequencies between 4 and 7 Hz are almost missing from the FWI adjoint fields, but are present in the VAFWI adjoint fields, and there is a general broadening (especially for the trace at $x=0.5$ km) and a shift towards lower frequencies in the VAFWI adjoint field spectrum. Thus, different handling of the receiver ghosts by VAFWI and FWI results in better preservation of the low frequency content in the VAFWI adjoint fields in the presence of the free surface. We show in the following sections that the information contained in receiver ghosts can significantly improve the reconstruction quality when the data are severely spatially undersampled and aliased.

\section{Examples}

\subsection{VAFWI of spatially undersampled data}

In this section we show examples of VAFWI performance with various receiver placements, including scenarios when the data are severely spatially undersampled and aliased. We compare VAFWI with conventional FWI. In all examples of this subsection we assume constant density for both data generation and inversion. 

We invert a part of the modified Marmousi-II velocity model \citep{Martin:2002}, where the modification is to make the water column shallower and to downsample. The water column is not updated during the inversion. The true velocity model is shown in Figure \ref{fig:C_geop}. The initial model is obtained from the true model by Gaussian smoothing and is shown in Figure \ref{fig:C0_geop}. The model is 3 km wide and 1.4 km deep. It is discretized into 151 $\times$ 71 grid-points. We use a Ricker source wavelet with a peak frequency of 7 Hz, and the wavelet is assumed to be known. The sources and receivers are placed 0.1 km below the top of the model. In all examples, we use 8 sources uniformly distributed between 0.1 and 2.9 km in the horizontal direction, but we vary the number of receivers. We place absorbing boundary conditions on all sides of the model for data generation and inversion. We run the inversions until convergence, i.e. until the objective function descent between two consecutive iterations is small or the line-search fails to find a new decent direction. This stopping strategy is not optimal for minimizing the computational cost, but we choose it over a predefined number of iterations in order to compare the best results obtainable by each method from the given starting model. 

\multiplot{2}{C_geop,C0_geop}{width=0.45\textwidth,trim={1cm 9cm 2cm 9cm},clip=true}
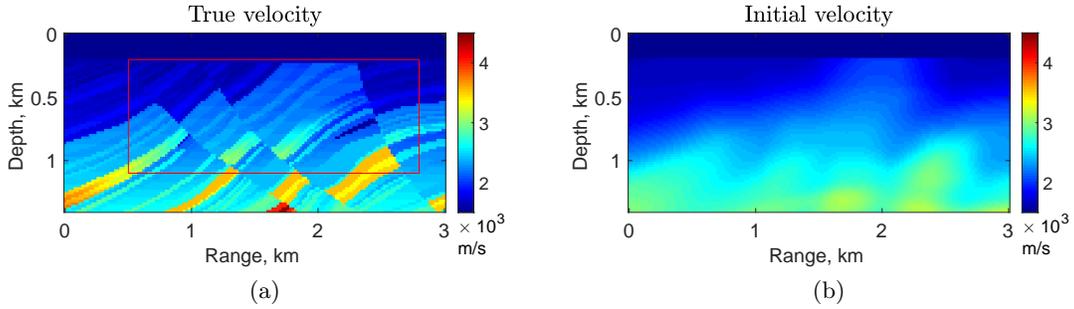
{True (a) and initial (b) Marmousi-II velocity model. The red box in (a) outlines the area where the recovered models are compared to the true model using equations (\ref{rmse}) and (\ref{ncc}).}

We quantify the quality of the recovered models using two measures of similarity to the true model:
\begin{itemize}
	\item Relative root mean square error between the true and the final recovered model:
		\begin{equation}
			RMSE = \frac{\sqrt{\sum_{x,z} \left[c_{rec}(x,z)-c_{true}(x,z)\right]^2}}{\sqrt{\sum_{x,z} c_{true}^2(x,z)}}\cdot 100 \%
			\label{rmse}
		\end{equation}
	\item Normalized cross-correlation between the true and the final recovered model:
		\begin{equation}
			NCC = \frac{\sum_{x,z} c_{rec}(x,z) \cdot c_{true}(x,z)}{\sqrt{\sum_{x,z} c_{rec}^2(x,z)} \sqrt{\sum_{x,z} c_{true}^2(x,z)}} \cdot 100 \% 
			\label{ncc}
		\end{equation}
\end{itemize}

We compute these measures on the part of the model between $x=0.5$ and $x=2.8$ km in the horizontal range, and between $z=0.2$ and $z=1.1$ km in depth. That is, we choose the part of the model where we can expect good model recovery based on reflector positions and survey geometry. This part of the model is shown as a red box in Figure \ref{fig:C_geop}. We also compute the value of the objective function (data misfit) in percentage of the initial misfit:
\begin{equation}
	\frac{J^{(end)}}{J^{(0)}}\cdot 100\%,
\end{equation} 
where $J^{(end)}$ and $J^{(0)}$ are final and starting objective values. In addition, we compare the cost for each inversion. This is done using the total number of LBFGS iterations as well as the total number of objective function and gradient evaluations (FG) per inversion run. As the inversion approaches the local minimum, the line-search algorithm tends to require more FGs per iteration. This significantly increases the cost of the inversion with little increment in recovered model accuracy. Therefore, the convergence criterion used by us in this subsection is not optimal from the computational point of view. Moreover, one needs to keep in mind that the shape of the objective functions for VAFWI and FWI may be different, affecting convergence behaviour and the computational cost of the two inversions. Therefore, the cost comparison may differ for a more optimal stopping criterion. The results of all the inversions are summarized in Table~\ref{tbl:undersampled}.

First, we run two benchmark examples of VAFWI and FWI with receivers placed at every grid-point between $x=0.1$ and 2.9 km in the horizontal direction. This receiver sampling is sufficiently dense to avoid data aliasing. The results of these tests are shown in Figures \ref{fig:C_vfwi_141r} and \ref{fig:C_fwi_141r}. The number of iterations for the two inversions is similar, but the number of FGs is significantly lower for VAFWI. The objective function reduction is similar and quite significant for both inversions. We can expect this, since the initial model is quite close to the the true model kinematically, and the inverted models are close to the global minimum. The RMSE and NCC for the recovered models are also very similar, and visually, the recovered models are indistinguishable from each other. FWI data is highly redundant in this case, and dense receiver placement ensures that information about the horizontal gradient of pressure is contained in the pressure data, offseting the benefit of using additional data components. 

Figure \ref{fig:C_vfwi_7r} shows VAFWI result with 7 receivers placed every 400 m between $x=0.3$ and 2.7 km. Figure \ref{fig:C_fwi_7r} shows the FWI result with the same receiver spacing. This sampling is sparse, causing data aliasing. For example, at the peak frequency of 7 Hz our sampling is about one-quarter of the Nyquist rate. Correspondingly, the recovered velocity models contain noise in the form of clutter. Quantitatively the VAFWI recovery is superior to FWI recovery, and visually VAFWI recovery is more coherent and less contaminated with clutter. The computational cost is comparable for the two inversions.

For comparison, Figure \ref{fig:C_fwi_13r} shows the FWI result with 13 pressure measurements equally spaced every 200 m between $x=0.3$ and 2.7 km. In this case, the receiver sampling is twice as dense as in the previous two examples. We note that quantitatively and qualitatively this reconstruction is very similar to the VAFWI recovery in Figure \ref{fig:C_vfwi_7r}. Thus with aliased data, one needs about twice the number of pressure data points to obtain recovery of the same quality as with vector data at a comparable computational cost. This is consistent with the conclusions of \cite{Robertsson:2008} for data interpolation. 
 
In some cases, the horizontal displacement component measurements may be unavailable or noisy. \cite{Ozbek:2010} show that the vertical component of velocity carries significant information in the horizontal direction, so we test how much reconstruction quality can be gained from only the vertical displacement component and pressure in VAFWI. Figure \ref{fig:C_vfwi_p_uz} shows VAFWI with 7 measurements of pressure $p$ and the vertical displacement component $u_z$. Qualitatively and quantitatively the recovery is noisier and less coherent than the VAFWI recovery with full vector data, but it is superior to the recovery with just pressure measurements (Figure \ref{fig:C_fwi_7r}).

\multiplot{6}{C_vfwi_141r,C_fwi_141r,C_vfwi_7r,C_fwi_7r,C_fwi_13r,C_vfwi_p_uz}
{width=0.45\textwidth,trim={1cm 9cm 2cm 8.5cm},clip=true}
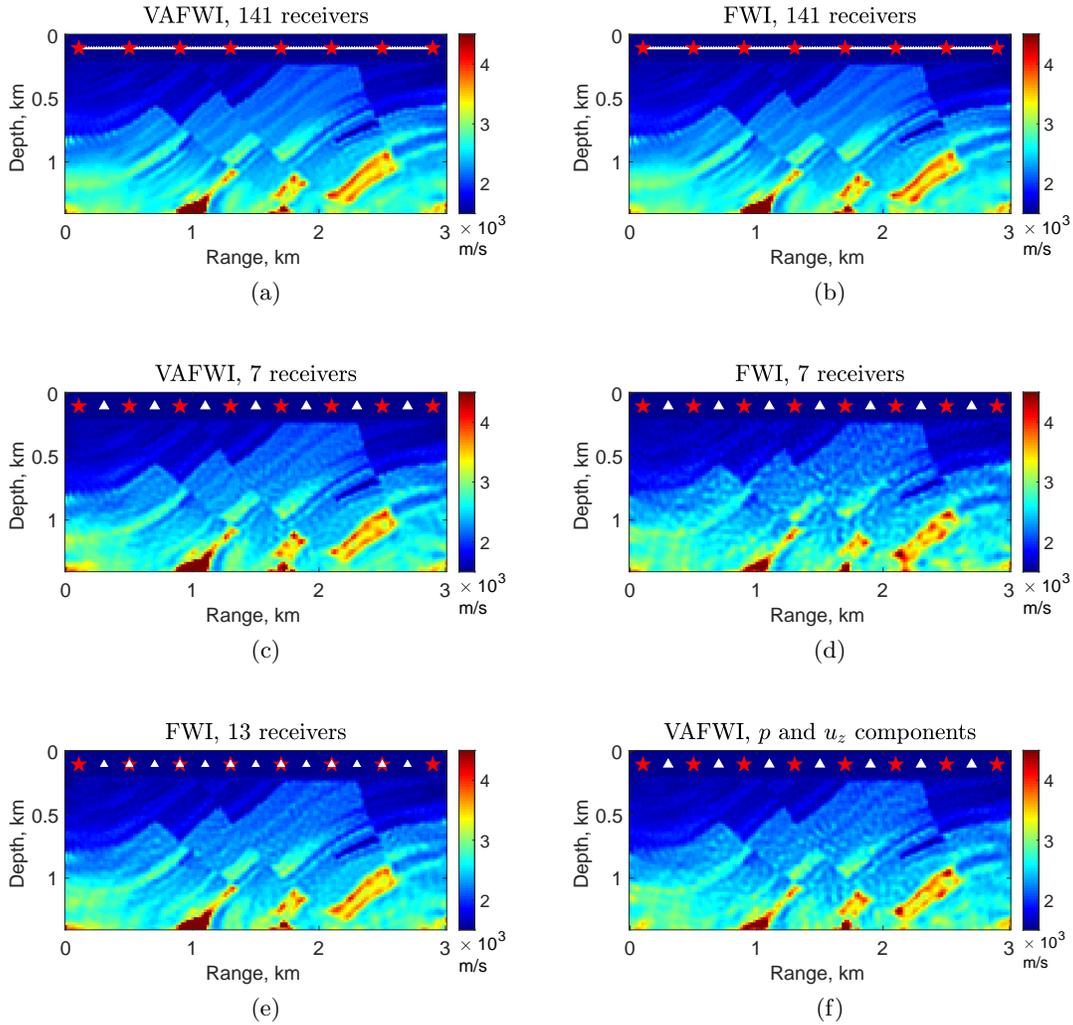
{Marmousi-II model reconstructions: (a) and (b) VAFWI and conventional FWI with dense receiver placement; receivers are placed at every grid point, there is no spatial data aliasing; (c) and (d) VAFWI and FWI with 7 receivers placed every 400 m, sparse receiver placement causes spatial data aliasing; (e) FWI with 13 receivers placed every 200 m; (f) VAFWI with receivers every 400 m, with $p$ and $u_z$ components only. Notation: red stars -- sources, white triangles -- receivers.}

\tabl{undersampled}{Performance of VAFWI and conventional FWI with various receiver placements.}{
  	\begin{center}
    	\begin{tabular}{|l|c|c|c|c|c|c|}
      		\hline
      		Inversion & Fig. & \# iter. & \# FGs & $J^{(end)}/J^{(0)}$ & RMSE, \% & NCC, \% \\
		\hline
      		\hline
     			VAFWI, 141 receivers & \ref{fig:C_vfwi_141r} & 99 & 155 & 0.07\% & 4.32 & 99.91 \\
      		\hline
     			FWI, 141 receivers & \ref{fig:C_fwi_141r} & 104 & 235 & 0.08\% & 4.36 & 99.91 \\
      		\hline
      			VFWI, 7 receivers & \ref{fig:C_vfwi_7r} & 154  & 379 & 0.03\% & 4.67 & 99.89 \\
      		\hline
      			FWI, 7 receivers & \ref{fig:C_fwi_7r} & 195 & 354 & 0.02\% & 6.12 & 99.81 \\
      		\hline
			FWI, 13 receivers & \ref{fig:C_fwi_13r} & 164 & 319 & 0.03\% & 4.68 & 99.89\\
      		\hline
			VFWI, 7 recevers, & \ref{fig:C_vfwi_p_uz} & 134 & 201 & 0.04\% & 5.55 & 99.85 \\
			$p, u_z$ components only & & & & & & \\
		\hline
    	\end{tabular}
  	\end{center}
}

\subsection{VAFWI of spatially undersampled data in the presence of free surface}

In this section we repeat some of the experiments from the previous subsection to show the effect of the free surface on velocity reconstruction from spacially undersampled data. The free surface boundary condition is placed at the top of the model for data generation and inversion. We run inversions until convergence and note the differences in the behaviour of the VAFWI and FWI objective functions. The results are summarized in Table~\ref{tbl:undersampled_fs}.

Figure \ref{fig:C_vfwi_141r_fs} shows the benchmark VAFWI reconstruction with receiver placement at every grid point between $x=0.1$ and 2.9 km in the horizontal direction. The FWI reconstruction for this benchmark example is visually the same and is not shown. Qualitatively and quantitatively both reconstructions are about as good as the one in Figure \ref{fig:C_vfwi_141r}. FWI took more iterations and FGs than VAFWI but it also reduced the objective function to a smaller percentage of its initial value. 

Figures \ref{fig:C_vfwi_7r_fs} and \ref{fig:C_fwi_7r_fs} show VAFWI and FWI recovery with 7 receivers spaced by 400 m. VAFWI recovery is almost as good as recovery with dense receiver coverage, meaning that the receiver ghosts helped fill in the missing data and further reduce the artifacts from data aliasing. In FWI, the free surface helped as well, but this recovery is somewhat noisier than the VAFWI result, and the bottom part of the model, where we did not measure the error, is visually less coherent. FWI took a lot longer to converge, but also reduced its objective significantly more. 

We now look more closely at the convergence behaviour of the two inversions. Figures \ref{fig:obj_fs} and \ref{fig:fg_fs} show objective function decay and the cumulative FGs count with iteration for the four inversions presented in this section. We note that the FGs count for all inversions initially grows linearly with iteration ($\sim$ 1 FG per iteration), and then increases more rapidly at the end of the inversion run. This happens because LBFGS tends to require more line searches per iteration as the inversion approaches the local minimum. Until approximately iteration 70, the objective function decay is about the same for the four inversions. By iteration 70 they require 72, 78, 74 and 75 FGs respectively and reduce the objective function to 0.06-0.07\% of its starting value. At the later iterations, the objective function decay and the computational cost curves diverge. In particular, VAFWI reaches its local minima in fewer iterations in both inversion runs than FWI, but these local minima appear to be slightly farther from the global minima than for FWI. We hypothesize that the differences in convergence behaviour may be due to the differences in frequency content of the adjoint fields or to other artifacts related to the free surface, but this hypotesis requires further investigation.

For comparison, Figures~\ref{fig:C_vfwi_7r_fs_70it} and \ref{fig:C_fwi_7r_fs_70it} show VAFWI and FWI results for the same experiment with undersampled data as in Figures \ref{fig:C_vfwi_7r_fs} and \ref{fig:C_fwi_7r_fs}, but after 70 iterations. Now the computational cost is comparable, however, the FWI result has degraded significantly more than the VAFWI result: visually, the faults are harder to delineate, and the bottom part of the model almost lacks coherence. The RMSE for these  reconstructions are 5.13 \% and 6.09 \%, and NCC are 99.87 \% and 99.81 \% respectively for VAFWI and FWI.

\multiplot{3}{C_vfwi_141r_fs,C_vfwi_7r_fs,C_fwi_7r_fs}{width=0.45\textwidth,trim={1cm 9cm 2cm 8.5cm},clip=true}
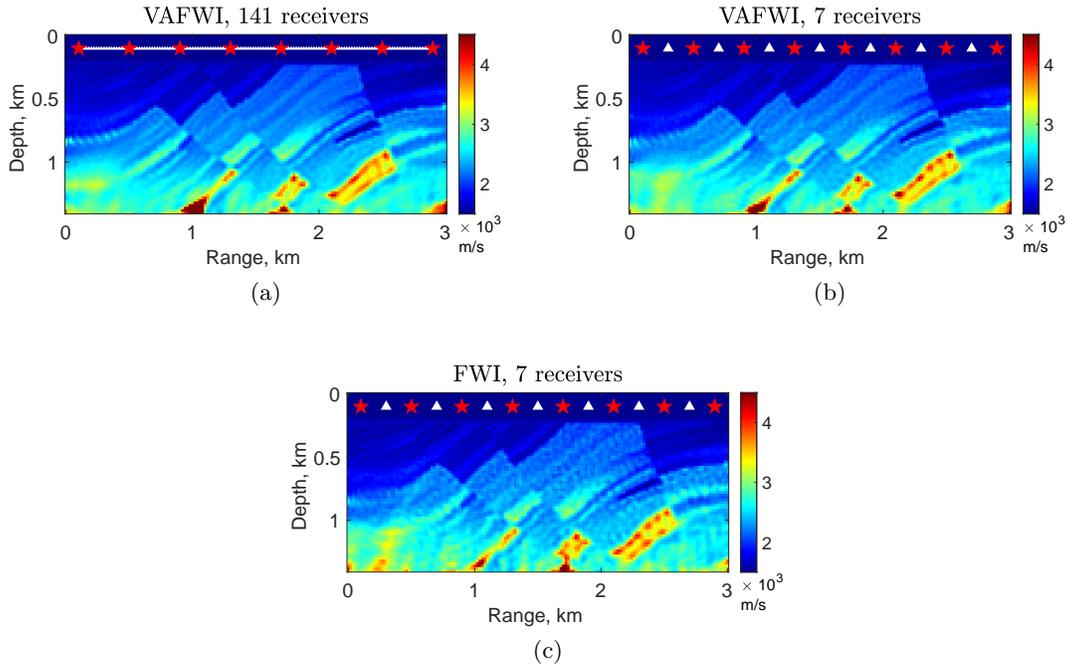
{Marmousi-II model reconstructions with free surface: (a) VAFWI with receivers every 20 m; (b) VAFWI with receivers every 400 m; (c) FWI with receivers every 400 m. Notation: red stars -- sources, triangles -- receivers.}

\multiplot{2}{obj_fs,fg_fs}{width=0.45\textwidth,trim={1cm 6cm 2cm 6cm},clip=true}
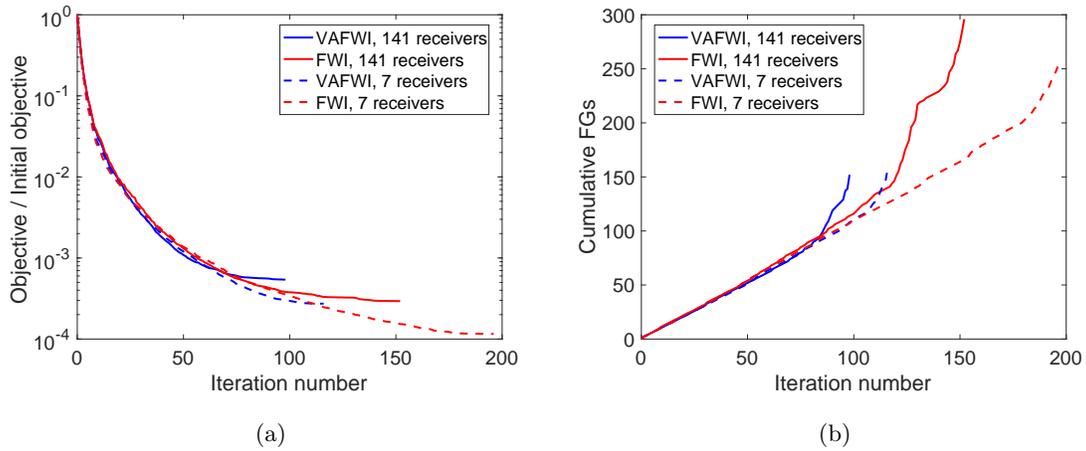
{(a) Objective function decay as a function of iteration number for the four examples in this section (Figure~\ref{fig:C_vfwi_141r_fs,C_vfwi_7r_fs,C_fwi_7r_fs}, Table~\ref{tbl:undersampled_fs}); (b) Cumulative FGs with iteration for the four examples.}

\tabl{undersampled_fs}{Performance of VAFWI and conventional FWI with various receiver placements in the presence of free surface.}{
  	\begin{center}
    	\begin{tabular}{|l|c|c|c|c|c|c|}
      		\hline
      		Inversion & Fig. & \# iter. & \# FGs & $J^{(end)}/J^{(0)}$ & RMSE, \% & NCC, \% \\
		\hline
      		\hline
     			VAFWI, 141 receivers & \ref{fig:C_vfwi_141r_fs} & 99 & 162 & 0.05\% & 4.56 & 99.90 \\
      		\hline
     			FWI, 141 receivers & Not shown & 152 & 296 & 0.03\% & 4.44 & 99.90 \\
      		\hline
      			VFWI, 7 receivers & \ref{fig:C_vfwi_7r_fs} & 116  & 158 & 0.03\% & 4.60 & 99.89 \\
      		\hline
      			FWI, 7 receivers & \ref{fig:C_fwi_7r_fs} & 196 & 252 & 0.01\% & 5.00 & 99.87 \\
      		\hline
    	\end{tabular}
  	\end{center}
}

\multiplot{2}{C_vfwi_7r_fs_70it,C_fwi_7r_fs_70it}{width=0.45\textwidth,trim={1cm 9cm 2cm 8.5cm},clip=true}
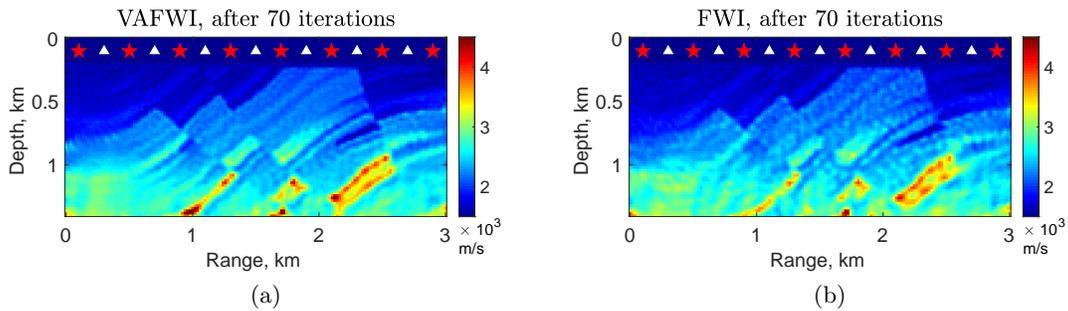
{Marmousi-II model reconstructions with free surface, after 70 iterations: (a) VAFWI with receivers every 400 m; (c) FWI with receivers every 400 m. Notation: red stars -- sources, triangles -- receivers.}

\subsection{Multi-parameter inversion}

The purpose of this section is to demonstrate that the de-aliasing property of vector data extends to multi-parameter inversion as well. We invert for P-wave velocity and density of the modified Marmousi-II model. The true and initial P-wave velocity models are the same as before (Figure~\ref{fig:C_geop,C0_geop}), and the true and initial density are shown in Figures~\ref{fig:rho} and \ref{fig:rho0}. Figure \ref{fig:Ip} shows the true acoustic impedance. The initial density model is obtained from the true density model by Gaussian smoothing. We place absorbing boundary conditions on all sides of the model. We use the same survey geometries as in the previous subsections and invert for velocity and density. We compare FWI and VAFWI results in the absence and presence of spatial data aliasing. All VAFWI inversion runs use full vector data. To keep the computational cost lower, we run 300 iterations of LBFGS for all inversions.

Figures~\ref{fig:C_vfwi_141r_c_rho} and~\ref{fig:rho_vfwi_141r_c_rho} show velocity and density recovered by VAFWI with dense receiver placement. The FWI recoveries are visually the same and are not shown to save space. The quantitative results for the two inversions are shown in the first two lines of Table~\ref{tbl:multi_models} and they are comparable. We note that long wavenumber components of velocity are well-recovered, while the short wavenumber components are slightly underestimated, whereas for density primarily the shorter wavenumber components are picked up by the inversion, while long wavenumber components are underemphasized, which is consistent with the radiation patterns analysis for the two parameters \citep{Forgues:1997}. We also obtain an accurate impedance reconstruction from the two models (not shown) as suggested by \cite{Operto:2013}. 

Figures \ref{fig:rho_vfwi_7r_c_rho}, \ref{fig:rho_fwi_7r_c_rho} show density models obtained by multi-parameter VAFWI and FWI with sparse receiver placement. We do not show recovered velocity models because the noise patterns in them are similar to the mono-parameter case. Figures \ref{fig:Ip_vfwi_7r_c_rho} and \ref{fig:Ip_fwi_7r_c_rho} show impedances obtained from the recovered velocities and densities by the two inversions. Quantitative results are shown in the last two rows of Table~\ref{tbl:multi_models}. As in the mono-parameter case, the VAFWI helps reduce noise from data aliasing in the recovered velocity and density. The resulting impedance model is also less noisy and more coherent. The cost of the two inversions and behaviour of objective functions are comparable, at least out to the 300 iterations we ran LBFGS for.

\multiplot{3}{rho,rho0,Ip}{width=.45\textwidth,trim={1cm 9cm 2cm 8.5cm},clip=true}
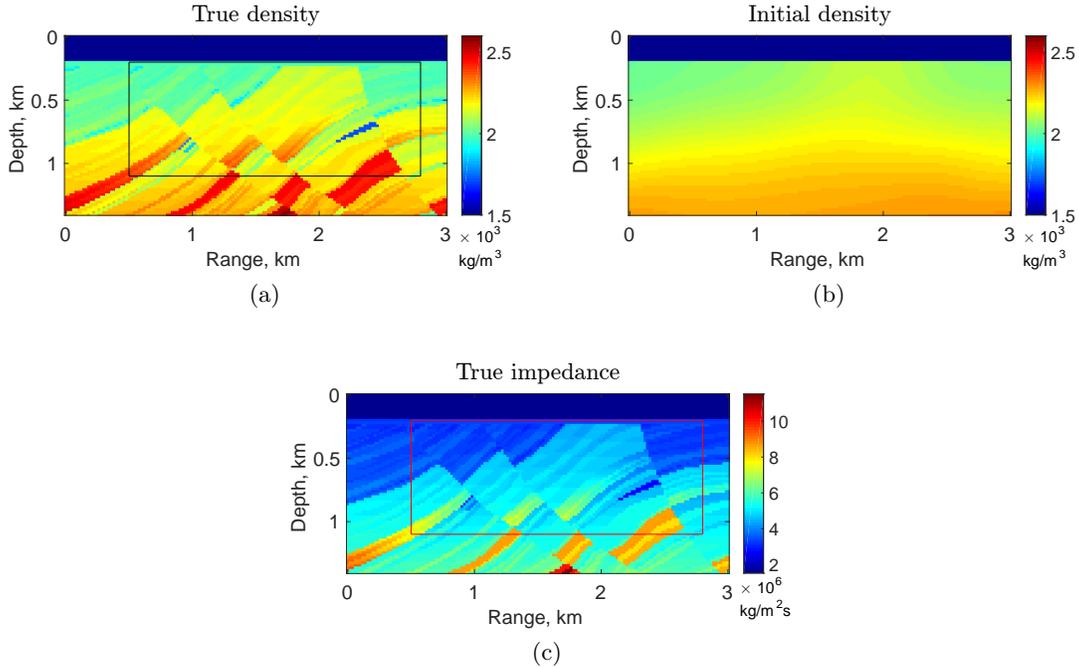
{(a) True density, (b) Initial density, (c) True impedance for multiparameter inversion. The boxes in images (a) and (c) outline the part of the model where measures (\ref{rmse}) and (\ref{ncc}) are calculated.}

\multiplot{6}{C_vfwi_141r_c_rho,rho_vfwi_141r_c_rho,rho_vfwi_7r_c_rho,rho_fwi_7r_c_rho,Ip_vfwi_7r_c_rho,Ip_fwi_7r_c_rho}{width=.45\textwidth,trim={1cm 9cm 2cm 8.5cm},clip=true}
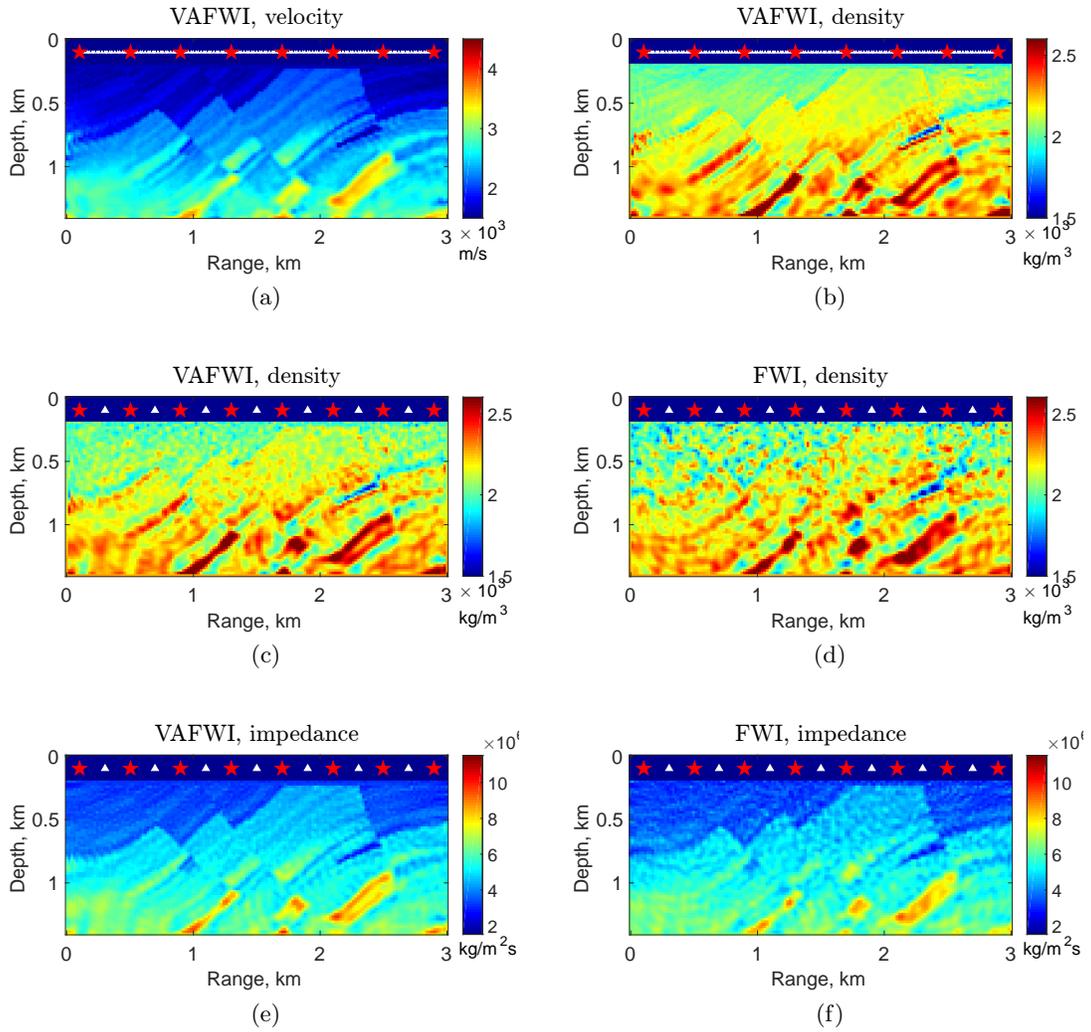
{Multi-parameter inversion: (a), (b) Velocity and density recovered by VAFWI with dense receiver placement; (c), (d) Density recovered by VAFWI and FWI with sparse receiver placement; (e), (f) Impedance models obtained from velocities and densities recovered by VAFWI and FWI with sparse receiver placement.}

\tabl{multi_models}{Performance of multi-parameter VAFWI and FWI: model recovery.}{
  	\begin{center}
    	\begin{tabular}{|l|c|c|c|c|c|c|c|c|c|}
      		\hline
      		\multirow{2}{*}{Inversion} & \multirow{2}{*}{Fig.} & \multicolumn{3}{|c|}{RMSE, \%} & \multicolumn{3}{|c|}{NCC, \%} & \multirow{2}{*}{$\frac{J^{(end)}}{J^{(0)}}$} & \multirow{2}{*}{\# FGs} \\
		\cline{3-8}
		& & $c$ & $\rho$ & $I_p$ & $c$ & $\rho$ & $I_p$ & & \\
		\hline
      		\hline
     			VAFWI, 141 rec. & $c, \rho$: \ref{fig:C_vfwi_141r_c_rho}, \ref{fig:rho_vfwi_141r_c_rho}  & 4.85 & 3.68 & 6.47& 99.88 & 99.93 & 99.79 & 0.05\% & 396 \\
      		\hline
		     	FWI, 141 rec. & Not shown & 5.00 & 3.63 & 6.68 & 99.88 & 99.93 & 99.78 & 0.05\% & 345 \\      		
		\hline
		     	VAFWI, 7 rec. & $\rho, I_p$: \ref{fig:rho_vfwi_7r_c_rho}, \ref{fig:Ip_vfwi_7r_c_rho} & 5.81 & 4.19 & 7.71 & 99.83 & 99.91 & 99.70 & 0.04\% & 327 \\
      		\hline
		     	FWI, 7 rec. & $\rho, I_p$: \ref{fig:rho_fwi_7r_c_rho}, \ref{fig:Ip_fwi_7r_c_rho} & 7.55 & 5.68 & 10.35 & 99.71 & 99.84 & 99.46 & 0.05\% & 336 \\    
		\hline
    	\end{tabular}
  	\end{center}
}

Multistage / multiscale inversion strategies can be used to further improve the reconstruction quality. We avoid them in this section on purpose, to make comparison clear and simple. We also note that the simplest and the most elegant formulas for the gradient update for multi-parameter VAFWI result when the model is parametrized in terms of compressibility $\kappa = \frac{1}{\rho c^2}$ and density $\rho$. In this case, the forward modelling equations depend linearly on the model parameters, and the gradients w.r.t. the model parameters decouple: $\frac{\partial J}{\partial \kappa}$ depends only on $p$ and $p^{\dagger}$, whereas $\frac{\partial J}{\partial \rho}$ depend only on $\ub$ and $\ub^{\dagger}$ \citep{Zheglova:2018}. \cite{Jeong:2012} propose a two-stage inversion strategy, where they invert for bulk modulus (the inverse of compressibility) at the first stage, followed by inversion of velocity and density at the second stage. They use the chain rule to compute the gradient update at the second stage. In our experience, an inversion workflow, in which compressibility is inverted first, followed by joint velocity and density inversion without the use of chain rule is slightly faster due to fast convergence at the first stage, but it also overestimates the short wavenumber components of the velocity model at the expense of those components in the density model. Quantitative improvements in the model recovery with this two-stage workflow were not significant in our experiments. Whether in certain situations the decoupling of the wavefields in the gradient components may help improve parameter separation and reduce the ill-posedeness of the VAFWI problem remains a goal for a future study.

\section{Conclusions}

In this paper, we described the vector acoustic full waveform inversion method first introduced by \cite{Akrami:2017} and extended to multi-parameter inversion by \cite{Zheglova:2018}. We showed that under specialized conditions the VAFWI adjoint operator becomes an inverse wavefield extrapolation operator, whereas the conventional FWI adjoint does not. This has implications for the way the two adjoints handle reflections off the free surface, particularly, receiver ghosts: the receiver ghost is separated from the reflected arrival in the VAFWI adjoint field, which results in constructive interference of the ghost with the reflected arrival. This leads to better preservation of low frequencies and fewer artifacts in the VAFWI adjoint field in the presence of the free surface and receiver ghosts. We compared performance of VAFWI and conventional FWI in situations when data are spatially undersampled causing data aliasing. We conclude that the de-aliasing property of vector data carries over to VAFWI, reducing aliasing artifacts in the reconstructed velocity, density and impedance models. Information contained in properly handled receiver ghosts helps further reduce these artifacts. These findings may be potentially useful in 3D inversion applications, where cross-line data sampling is often below the Nyquist rate or in other situations, when data are undersampled. Identifying the situations, in which VAFWI may result in better parameter separation and improvement of the ill-posedeness of the multi-parameter inverse problem remains a future goal, as is field data application.

\newpage

\bibliographystyle{seg}  
\bibliography{SEGrefs}

\begin{thebibliography}{}
\itemsep0pt

\bibitem[Aki and Richards, 2002]{Aki:2002}
Aki, K., and P.~G. Richards,  2002, Quantitative seismology: University Science
  Books.

\bibitem[Akrami et~al., 2017]{Akrami:2017}
Akrami, S., P. Zheglova, and A. Malcolm,  2017, An algorithm for vector data
  full-waveform inversion: SEG Technical Program Expanded Abstracts 2017,
  1568--1572.

\bibitem[Asnaashari et~al., 2012]{Asnaashari:2012}
Asnaashari, A., R. Brossier, C. Castellanos, B. Dupuy, V. Etienne, Y. Gholami,
  G. Hu, L. Metivier, S. Operto, D. Pageot, V. Prieux, A. Ribodetti, A. Roques,
  and J. Virieux,  2012, Hierarchical approach of seismic full waveform
  inversion: Numerical Analysis and Applications, {\bf 5}, 99--108.

\bibitem[Brossier et~al., 2009]{Brossier:2009}
Brossier, R., S. Operto, and J. Virieux,  2009, Seismic imaging of complex
  onshore structures by 2d elastic frequency-domain full-waveform inversion:
  Geophysics, {\bf 74}, WCC105--WCC118.

\bibitem[Carlson et~al., 2007]{Carlson:2007}
Carlson, D., A. Long, W. S{\"o}llner, H. Tabti, R. Tenghamn, and N. Lunde,
  2007, Increased resolution and penetration from a towed dual‐sensor
  streamer: First break, {\bf 25}, 71--77.

\bibitem[Cassereau and Fink, 1992]{Cassereau:1992}
Cassereau, D., and M. Fink,  1992, Time-reversal of ultrasonic fields. iii.
  theory of the closed time-reversal cavity: IEEE Transactions on Ultrasonics,
  Ferroelectrics, and Frequency Control, {\bf 39}, 579--592.

\bibitem[Choi et~al., 2008]{Choi:2008}
Choi, Y., D.-J. Min, and C. Shin,  2008, Two-dimensional waveform inversion of
  multi-component data in acoustic-elastic coupled media: Geophysical
  Prospecting, {\bf 56}, 863–881.

\bibitem[Fichtner, 2011]{Fichtner:2011}
Fichtner, A.,  2011, Full seismic waveform modelling and inversion: Springer.

\bibitem[Fleury and Vasconcelos, 2013]{Fleury:2013}
Fleury, C., and I. Vasconcelos,  2013, Adjoint-state reverse time migration of
  4c data: Finite-frequency map migration for marine seismic image: Geophysical
  Journal International, {\bf 78}, WA159--WA172.

\bibitem[Forgues and Lambar{\'e}, 1997]{Forgues:1997}
Forgues, E., and G. Lambar{\'e},  1997, Parametrization study for acoustic and
  elastic ray + born inversion: Journal of Seismic Exploration, {\bf 6},
  253--277.

\bibitem[Halliday et~al., 2012]{Robertsson:2012}
Halliday, D., J.~O.~A. Robertsson, I. Vasconcelos, D.-J. van Manen, R. Laws, K.
  Özdemir, and H. Grønaas,  2012, Full-wavefield, towed-marine seismic
  acquisition and applications: SEG Technical Program Expanded Abstracts 2012.

\bibitem[Jeong and Min, 2012]{Jeong:2012}
Jeong, W., and D.-J. Min,  2012, Application of acoustic full waveform
  inversion for density estimation: SEG Technical Program Expanded Abstracts
  2012.

\bibitem[Kohnke and Sava, 2019]{Kohnke:2019}
Kohnke, C., and P. Sava,  2019, Inversion of vector-acoustic data in a local
  domain: SEG Technical Program Expanded Abstracts.

\bibitem[Martin et~al., 2002]{Martin:2002}
Martin, G.~S., K.~J. Marfurt, and S. Larsen,  2002, Marmousi‐2: An updated
  model for the investigation of avo in structurally complex areas: SEG
  Technical Program Expanded Abstracts,  1979--1982.

\bibitem[Meier et~al., 2015]{Meier:2010}
Meier, M.~A., R.~E. Duren, K.~T. Lewallen, J. Otero, S. Heiney, and T. Murray,
  2015, A marine dipole source for low frequency seismic acquisition: SEG
  Technical Program Expanded Abstracts.

\bibitem[Morse and Feshbach, 1953]{Morse:1953}
Morse, P.~M., and H. Feshbach,  1953, Methods of theoretical physics:
  McGraw-Hill.

\bibitem[Nocedal and Wright, 2006]{Nocedal:2006}
Nocedal, J., and S.~J. Wright,  2006, Numerical optimization: Springer.

\bibitem[Operto et~al., 2013]{Operto:2013}
Operto, S., Y. Gholami, V. Prieux, A. Ribodetti, R. Brossier, L. Metivier, and
  J. Virieux,  2013, A guided tour of multiparameter full-waveform inversion
  with multicomponent data: From theory to practice: The Leading Edge, {\bf
  32}, 1040--1054.

\bibitem[Orji, 2012]{Orji:2012}
Orji, O.~C.,  2012, Sea surface wave height estimation from dual-sensor towed
  streamer: PhD thesis, University of Oslo.

\bibitem[{\"O}zbek et~al., 2010]{Ozbek:2010}
{\"O}zbek, A., M. Vassallo, K. {\"O}zdemir, D.-J. van Manen, and K.
  Eggenberger,  2010, Crossline wavefield reconstruction from multicomponent
  streamer data: Part 2 - joint interpolation and 3d up/down separation by
  generalized matching pursuit: Geophisics, {\bf 75}, WB69--WB85.

\bibitem[Plessix, 2006]{Plessix:2006}
Plessix, R.-E.,  2006, A review of the adjoint-state method for computing the
  gradient of a functional with geophysical applications: Geophysical Journal
  International, {\bf 167}, 495--503.

\bibitem[Plessix et~al., 2013]{Plessix:2013}
Plessix, R.~E., P. Milcik, H. Rynja, A. Stopin, K. Matson, and S. Abri,  2013,
  Multiparameter full-waveform inversion: Marine and land examples: The Leading
  Edge, {\bf 32}, 1030--1038.

\bibitem[Prieux et~al., 2013]{Prieux:2013}
Prieux, V., R. Brossier, S. Operto, and J. Virieux,  2013, Multiparameter full
  waveform inversion of multicomponent ocean-bottom-cable data from the valhall
  field. part 1: imaging compressional wave speed, density and attenuation:
  Geophysical Journal International, {\bf 194}, 1640--1664.

\bibitem[Ravasi et~al., 2015]{Ravasi:2015}
Ravasi, M., I. Vasconcelos, A. Curtis, and A. Kritski,  2015, Vector-acoustic
  reverse time migration of volve ocean-bottom cable data set without up/down
  decomposed wavefields: Geophysics, {\bf 80}, S137--S150.

\bibitem[Reiser et~al., 2012]{Reiser:2012}
Reiser, C., T.~L. Bird, F. Engelmark, E.~C. Anderson, and Y. Balabekov,  2012,
  Value of broadband seismic for interpretation, reservoir characterization and
  quantitative interpretation workflows: First Break, {\bf 30}, 67--75.

\bibitem[Robertsson et~al., 2008]{Robertsson:2008}
Robertsson, J. O.~A., I. Moore, M. Vassallo, K. {\"O}zdemir, D.-J. {v}an Manen,
  and A. {\"O}zbek,  2008, On the use of multicomponent streamer recordings for
  reconstruction of pressure wavefields in the crossline direction: Journal of
  Geophysical Research, {\bf 73}, A45–A49.

\bibitem[S{\"o}llner et~al., 2008]{Soellner:2008}
S{\"o}llner, W., E. Brox, M. Widmaier, and S. Vaage,  2008, Surface related
  multiple suppression in dual-sensor towed streamer data: 70th EAGE Conference
  and Exhibition incorporating SPE EUROPEC 2008.

\bibitem[Vassallo et~al., 2010]{Vassallo:2010}
Vassallo, M., A. {\"O}zbek, K. {\"O}zdemir, and K. Eggenberger,  2010,
  Crossline wavefield reconstruction from multicomponent streamer data: Part 1
  — multichannel interpolation by matching pursuit (mimap) using pressure and
  its crossline gradient: Geophisics, {\bf 75}, WB53--WB67.

\bibitem[Wapenaar, 2007]{Wapenaar:2007}
Wapenaar, K.,  2007, General representations for wavefield modeling and
  inversion in geophysics: Geophisics, {\bf 72}, SM5--SM17.

\bibitem[Wapenaar and Fokkema, 2004]{Wapenaar:2004}
Wapenaar, K., and J. Fokkema,  2004, Reciprocity theorems for diffusion, flow
  and waves: Journal of Applied Mechanics, {\bf 71}, 145–150.

\bibitem[Yang et~al., 2014]{Yang:2014}
Yang, J., Y. Liu, and L. Dong,  2014, A multi-parameter full waveform inversion
  strategy in acoustic media: 76th EAGE Conference and Exhibition Expanded
  Abstracts.

\bibitem[Yang and Malcolm, 2019]{Yang:2019}
Yang, Q., and A. Malcolm,  2019, Single parameter full waveform inversion in
  fluid-saturated porous media: SEG Technical Program Expanded Abstracts 2019,
  accepted.

\bibitem[Yang et~al., 2018]{Yang:2018}
Yang, Q., A. Malcolm, H. Rusmanugroh, and W. Mao,  2018, Parameter tradeoffs in
  fwi in fluid-saturated porous media: SEG Technical Program Expanded Abstracts
  2018,  1334--1338.

\bibitem[Yilmaz, 2008]{Yilmaz:2008}
Yilmaz, {\"O}.,  2008, Seismic data analysis: Society of Exploration
  Geophysicists.

\bibitem[Zheglova et~al., 2018]{Zheglova:2018}
Zheglova, P., S.~M. Akrami, and A. Malcolm,  2018, Multi-parameter vector
  acoustic full waveform inversion: SEG Technical Program Expanded Abstracts
  2018,  1354--1358.

\bibitem[Zheglova and Malcolm, 2019]{Zheglova:2019}
Zheglova, P., and A. Malcolm,  2019, Taking advantage of receiver ghosts and
  pressure gradients in vector acoustic full waveform inversion: SEG Technical
  Program Expanded Abstracts 2019,  accepted.

\bibitem[Zhong and Liu, 2019]{Zhong:2019}
Zhong, Y., and Y. Liu,  2019, Source-independent time-domain vector-acoustic
  full-waveform inversion: Geophysics, {\bf 84}, R489–R505.

\end{thebibliography}

\end{document}